\documentclass[preprint, 5p, twocolumns,sort&compress]{elsarticle}

\usepackage{graphicx}
\usepackage{hyperref}
\usepackage{siunitx}
\usepackage{color}
\usepackage{amsmath}
\usepackage{lineno}

\title{Absolute X-ray energy measurement using a high-accuracy angle encoder}

\begin{document}                  % DO NOT DELETE THIS LINE
\begin{frontmatter}
     %-------------------------------------------------------------------------
     % The introductory (header) part of the paper
     %-------------------------------------------------------------------------

     % The title of the paper. Use \shorttitle to indicate an abbreviated title
     % for use in running heads (you will need to uncomment it).

%
%
\author[okayama]{Takahiko~Masuda\corref{firstfoot}}
\ead{masuda@okayama-u.ac.jp}
\author[aist]{Tsukasa~Watanabe\corref{firstfoot}}
\author[wien]{Kjeld~Beeks}
\author[aist]{Hiroyuki~Fujimoto}
%\author[okayama]{Hideaki~Hara}
\author[okayama]{Takahiro~Hiraki}
\author[okayama]{Hiroyuki~Kaino}
\author[kyoto]{Shinji~Kitao}
\author[okayama]{Yuki~Miyamoto}
\author[okayama]{Koichi~Okai}
\author[okayama]{Noboru~Sasao}
\author[kyoto]{Makoto~Seto}
\author[wien]{Thorsten~Schumm}
\author[riken]{Yudai~Shigekawa}
\author[sp8]{Kenji~Tamasaku}
\author[okayama]{Satoshi~Uetake}
\author[riken]{Atsushi~Yamaguchi}
\author[jasri]{Yoshitaka~Yoda}
\author[okayama]{Akihiro~Yoshimi}
\author[okayama]{Koji~Yoshimura}

\address[okayama]{Research Institute for Interdisciplinary Science, Okayama University, Okayama, Japan}
\address[aist]{National Institute of Advanced Industrial Science and Technology, Tsukuba, Japan}
\address[wien]{Institute for Atomic and Subatomic Physics - Atominstitut, TU Wien, Vienna, Austria}
\address[kyoto]{Institute for Integrated Radiation and Nuclear Science, Kyoto University, Kumatori-cho, Japan}
\address[riken]{RIKEN, Wako, Japan}
\address[sp8]{RIKEN, SPring-8 Center, Sayo-cho, Japan}
\address[jasri]{Japan Synchrotron Radiation Research Institute, Sayo-cho, Japan}

\cortext[firstfoot]{These authors contributed equally.}

\begin{abstract}
 This paper presents an absolute X-ray photon energy measurement method that uses a Bond diffractometer. The proposed system enables the prompt and rapid \textit{in-situ} measurement of photon energies in a wide energy range. The diffractometer uses a reference silicon single crystal plate and a highly accurate angle encoder called SelfA. We evaluate the performance of the system by repeatedly measuring the energy of the first excited state of the potassium-40 nuclide. The excitation energy is determined as \SI{29829.39\pm0.06}{eV}. It is one order of magnitude more precise than the previous measurement. The estimated uncertainty of the photon energy measurement was 0.7\,ppm as a standard deviation and the maximum observed deviation was 2\,ppm.
\end{abstract}

\end{frontmatter}
     %-------------------------------------------------------------------------
     % The main body of the paper
     %-------------------------------------------------------------------------
     % Now enter the text of the document in multiple \section's, \subsection's
     % and \subsubsection's as required.

%\linenumbers

\section{Introduction}

A synchrotron radiation X-ray beam is often monochromatized by silicon monochromators installed in the beam line. The energy bandwidth of this monochromatized beam is usually a few electron volts (eV), narrower than sub-eV bandwidth is also easily available by using high-resolution monochromators. The monochromatic beam is used in various research fields that require an accurate photon energy.
For example, in nuclear resonant scattering (NRS) experiments \cite{Seto2012}, narrow resonance peaks of NRS sometimes need to be found. Knowing the accurate photon energy helps in finding the resonance without wide energy scanning, which is especially useful for searching weak resonance peaks.
Accurate photon energy information is also needed to determine the lattice spacing of samples for the structural determination of crystalline, non-crystalline, and nano-materials \cite{Billinge2007}.

While various absolute energy calibration methods have been proposed thus far \cite{Arthur1989, Kraft1996,Hong2012}, the most commonly used technique is to measure an absorption edge of a reference element.
Although the absolute photon energy is adjusted at a certain absorption edge, the absolute photon energy easily deviates over time because of uncontrollable factors, such as a change in the thermal distribution of the monochromators. In particular, in NRS measurements, the photon energy is scanned to search for a resonance peak; the scanning operation changes the monochromator condition and energy drifts are unavoidable. The absorption edge measurement method cannot overcome such problems.
Therefore, an easily available \textit{in-situ} method that can calibrate the absolute photon energy in a wide energy range is highly appreciated.

The authors recently demonstrated the NRS measurement of the thorium-229 nucleus \cite{Masuda2019} to study its low energy first excited state, called an isomer \cite{Thielking2018,Burke2019,Seiferle2019}. In this work, the thorium nuclei were resonantly excited to the second nuclear state by irradiating them with a monochromatic X-ray beam. The accurate X-ray photon energy measurement was a key component of the work. Since the resonance is weak and there is a large non-resonant scattering background, the X-ray photon energy had to be sufficiently monochromatized and stabilized over a long period. More specifically, the resonance energy was \SI{29}{keV}, and the bandwidth of the X-ray photon energy was \(\sim\)\SI{0.1}{eV}; therefore, the photon energy had to be monitored at a 1\,ppm level during 24-hour scanning.
Further, the absolute energy can be used to accurately determine the isomer energy by combining with high-accuracy gamma-ray spectroscopies \cite{Yamaguchi2019}.
 
The absolute energy measurement system used in the present study utilizes a method developed by Bond \cite{Bond1960}, which was originally used for crystal lattice spacing measurements. 
This method is based on Bragg's law: the relation between the wavelength of an X-ray photon \(\lambda_\mathrm{Xray}\), which is the inverse of the photon energy \(E_\mathrm{Xray}\), and the crystal lattice spacing \(d\) is \(
 \lambda_\mathrm{Xray} = 2 d \sin \theta_\mathrm{B},
\)
where \(\theta_\mathrm{B}\) is the Bragg angle of the crystal. To increase the measurement accuracy, Bond proposed and designed a measurement scheme in which the X-ray Laue diffraction angle of a reference crystal is measured at both sides of the primary beam. This method can eliminate zero-point offset because it uses the angle difference between both diffractions. It can also eliminate possible setting errors of the crystal by \textit{in-situ} alignment. The Bond method has been widely used for the lattice spacing determination of various crystals, and the reported uncertainties are \(\Delta d/d \sim 10^{-5}\) \cite{Herbstein2000,Schmidbauer2012}.

For the X-ray photon energy calibration, the Bragg angle is measured with a reference crystal whose lattice spacing is well-known. The lattice spacing of a silicon single crystal has been well established with an accuracy of \(\Delta d/d \sim 10^{-7}\) \cite{Cavagnero2004,Cavagnero2004e}. Therefore, the key to an accurate absolute energy measurement is the accurate measurement of the diffraction angle. 
For example, the accurate energy calibration of X-ray absorption edges on the order of \(10^{-5}\) to \(10^{-6}\) has been reported by using a silicon reference crystal and a Bond diffractometer \cite{Kraft1996}; the angle uncertainty was \(\pm 0.12\)\,arcsecond measured by a dedicated calibration method \cite{Becker1990}.

In the present study, we used a rotary encoder equipped with a self-calibration function, called SelfA \cite{Watanabe2005} for an accurate angle measurement. It was originally developed in the National Metrology Institute of Japan and is now commercially available. The accuracy of the angle measurement is better than \SI{0.1}{arcsecond}. 

In this study, we adopted the accurate rotary encoder based on SelfA for a Bond diffractometer to the absolute photon energy measurement of the synchrotron radiation X-ray beam. This method can perform \textit{in-situ} measurements within \(\sim\)3 minutes and it can be applied to any energy range. The apparatus can be easily positioned at the downstream end of the experiment such that the main experiment is not disturbed. We took repeated measurements of the excitation energy of the first excited state of the potassium-40 (\(^{40}\)K) nucleus with various settings to verify the reliability of this method.

\section{Bond diffractometer} \label{sec:diffractometer}

\subsection{Principle}
\begin{figure*}
\includegraphics[width=18cm, bb=0 0 1100 750, clip]{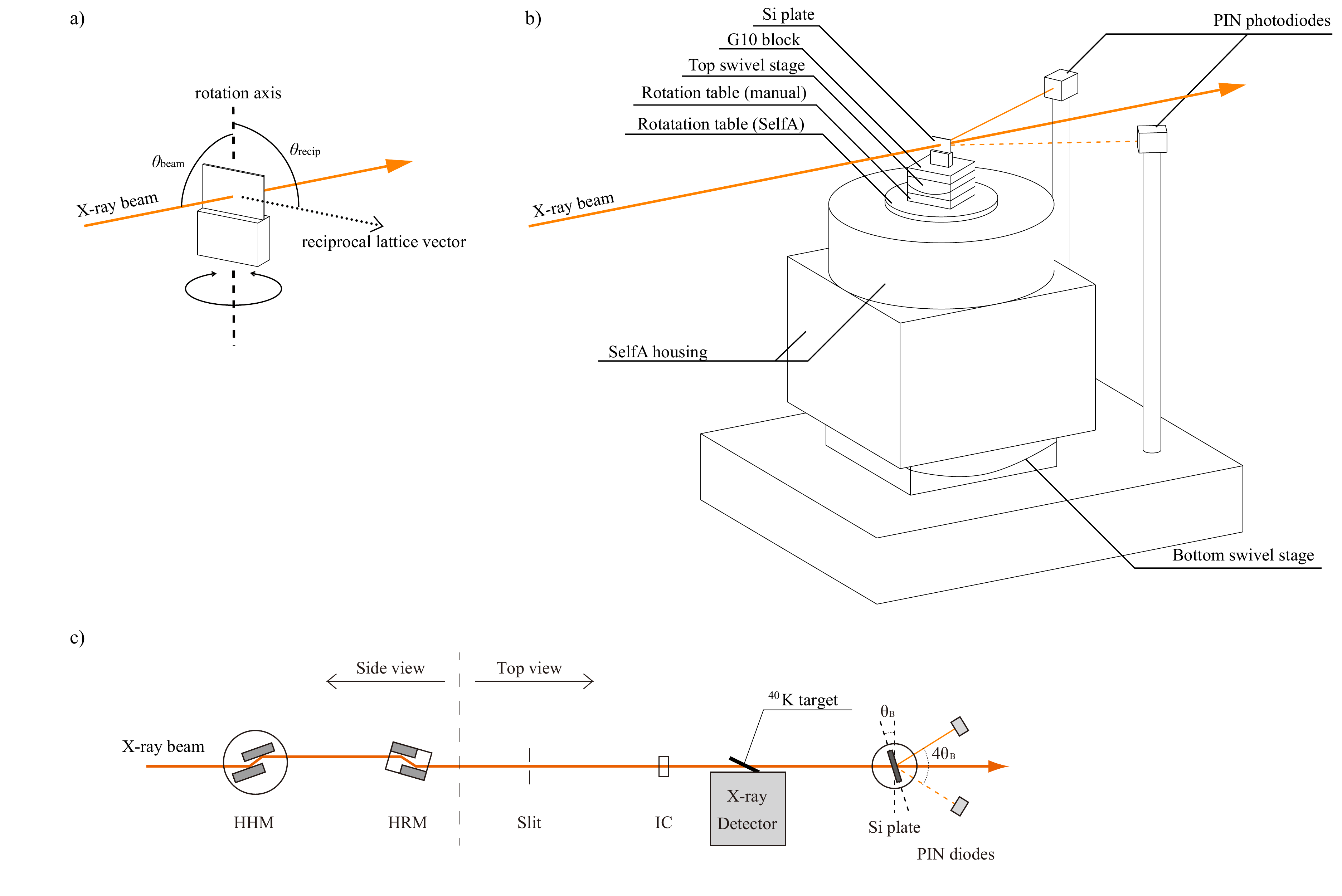}
\caption{a) Schematic of the Si plate. Geometric arrangement of the X-ray primary beam, plate rotation axis, and reciprocal lattice vector of the crystal is drawn. b) Schematic of the Bond diffractometer. c) Schematic of the beam line. HHM: High heat-load monochromator; HRM: High resolution monochromator; IC: Ionization chamber; \(\theta_\mathrm{B}\): Bragg angle. 
 The X-ray beam shown as the orange line comes in from the left side and passes through the Si plate in each panel. The diffracted beams are indicated by thin orange lines. In sub-figures b and c, the Si plate is rotated by the Bragg angle from the angle perpendicular to the beam, and one PIN photodiode receives the diffracted beam. }
\label{fig:system}
\end{figure*}

The Bond diffractometer measures the absolute energy of X-ray photons based on the diffraction angle (\(\theta_\mathrm{B}\)) from a silicon single crystal plate (Si plate).
The accurate relation between the X-ray photon energy \(E_\mathrm{Xray}\) and the diffraction angle is
\begin{align} 
 E_\mathrm{Xray} &= \frac{1.239841857~ \mathrm{[keV \cdot nm]} }{2 d(P,T) \sin \theta_\mathrm{B} \sin \theta_\mathrm{beam} \sin \theta_\mathrm{recip}}, \label{eq:conversion} \\
  d(P,T) &= \frac{d_{220}}{2} \left( 1-\frac{1}{3}PC_\mathrm{comp} \right) \notag \\ &~~~~~~~~~~~~~ \times \left( 1+ (T - 22.5[\mathrm{^\circ C}]) C_\mathrm{temp} \right), \label{eq:correction}
\end{align}
where \(P\) and \(T\) are ambient pressure and temperature, respectively, \(\theta_\mathrm{beam}\) is the angle between the Si plate rotation axis and the primary beam, and \(\theta_\mathrm{recip}\) is the angle between the Si plate rotation axis and the reciprocal lattice vector of the crystal. These angles are graphically shown in figure~\ref{fig:system}a. The numerator of Eq.~\ref{eq:conversion} is the conversion factor of the wavelength to the energy of X-ray photons. 
The lattice spacing \(d\) depends on the temperature and pressure as described in Eq.~\ref{eq:correction}. \(d_{220}\) is the lattice spacing of (220) lattice planes at \(T=22.5\)\,\si{\celsius} and in vacuum (\(P=0\,\mathrm{Pa}\)), \(C_\mathrm{temp}\) is a thermal expansion coefficient, and \(C_\mathrm{comp}\) is compressibility. The first 1/2 factor is added because we used (440) lattice planes.
The Si plate rotation axis has to be perpendicular to both the primary beam and the reciprocal lattice vector of the crystal so that \(\theta_\mathrm{beam} = \theta_\mathrm{recip}=\)\,\SI{90}{\degree} is satisfied. Deviations of these angles cause systematic uncertainty; their accuracy is not crucial however, for their first derivatives are zero.

\subsection{SelfA}\label{sec:selfa}
The diffraction angle is measured by a rotary encoder controlled with SelfA. SelfA utilizes the equal division averaged method \cite{Masuda1993} that relies on the fact that the angle is a \SI{360}{\degree} closed system. 
It analyzes the angular deviation by evaluating angular signals from multiple reading heads located at equal angular intervals around the grating lines on a rotary disk. The reading heads sense the grating lines that pass over the heads during a rotation. The signal detected from the grating lines contain the angular deviation from the ideal angular signal. 

If \textit{N} reading heads are arranged at \SI{360}{\degree}/\textit{N} intervals, the system can detect the angular deviation except for the $n \times N$-th order Fourier components (\textit{n} is an integer); therefore, the number of reading heads should be large to ensure accurate calibration \cite{Watanabe2014}.
Although a larger \textit{N} is required to improve calibration, which enables one to detect higher order Fourier components, the higher order components have less effect on the angle deviation than the lower order components.
Moreover, if other \textit{M} reading heads are arranged at \SI{360}{\degree}/\textit{M} intervals with the original \textit{N} reading heads, some $n \times N$-th order Fourier components can be obtained, and the remaining undetected Fourier components are only $n \times N \times M$-th order Fourier components. Therefore, the setup in which the \textit{N} reading heads arranged at \SI{360}{\degree}/\textit{N} intervals and \textit{M} reading heads arranged at \SI{360}{\degree}/\textit{M} intervals is equivalent to $N \times M$ reading heads arranged at \SI{360}{\degree}/$(N \times M)$ intervals. 

\begin{figure}
\includegraphics[width=7cm, bb=0 0 650 650, clip]{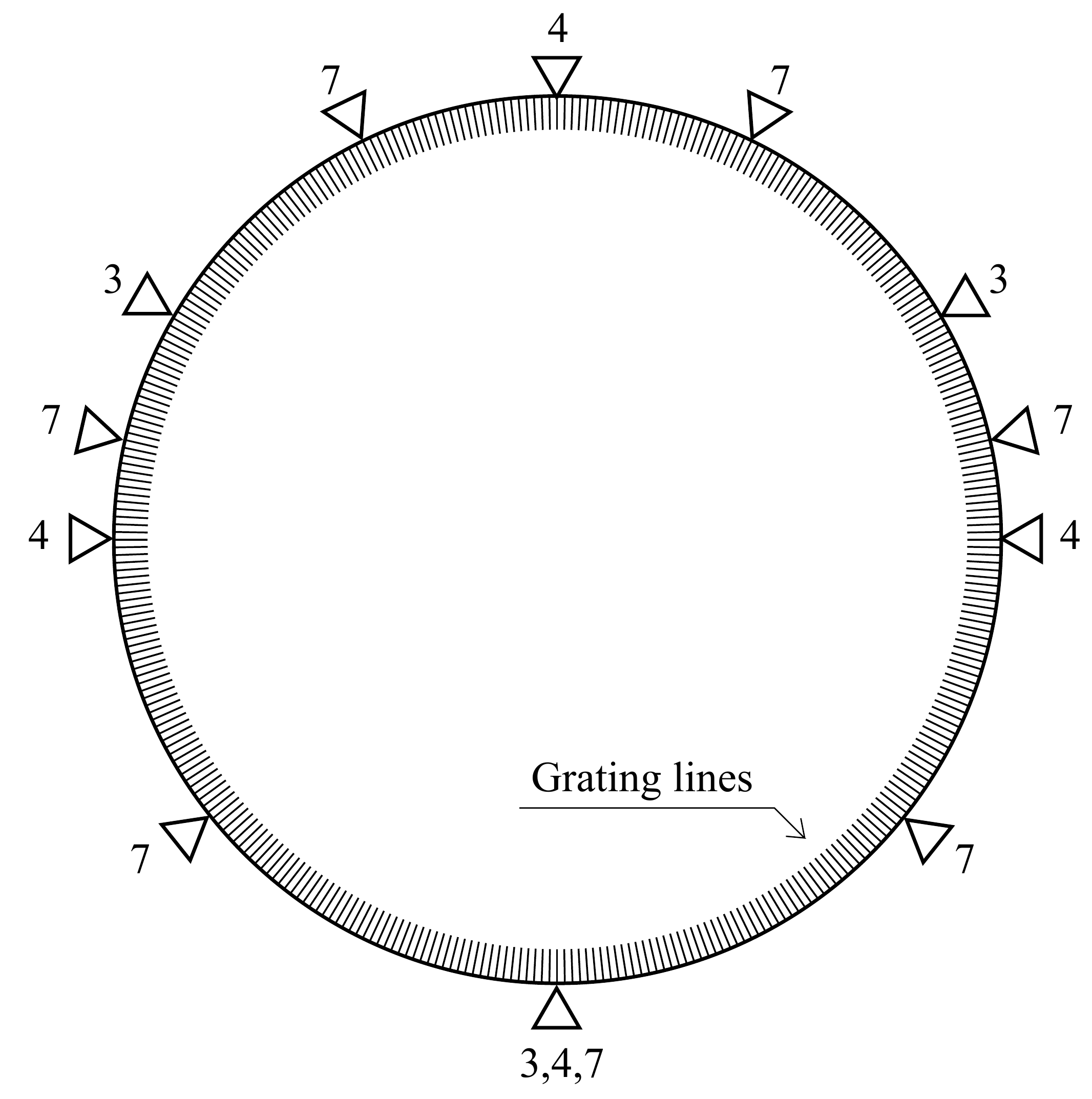}
\caption{Schematic of the rotary encoder equipped in SelfA. The twelve triangles represent the reading heads and their labels indicate their grouping.}
\label{fig:rotaryencoder}
\end{figure}

In this work, we used SelfA, which has twelve reading heads; these heads were placed at 1/3 (\SI{120}{\degree}), 1/4 (\SI{90}{\degree}), and 1/7 (\SI{51.43}{\degree}) positions around a rotary disk as shown in figure~\ref{fig:rotaryencoder}. This setup can be used to detect the angular deviation from the ideal exact equally spaced lines except for the 84\textit{n}-th order Fourier components.

The rotary encoder has 36000 grating lines with an angular pitch of \SI{0.01}{\degree}. These signals read by the heads are electrically divided into 1024 subpoints by an interpolating circuit between the grating pitches to increase the angular granularity to $36000 \times 1024$, which correspond to the angular pitch of \SI{0.03515625}{arcsecond}.

\subsection{Instruments}
\begin{figure}
\includegraphics[width=7cm, bb=200 0 2700 3024, clip]{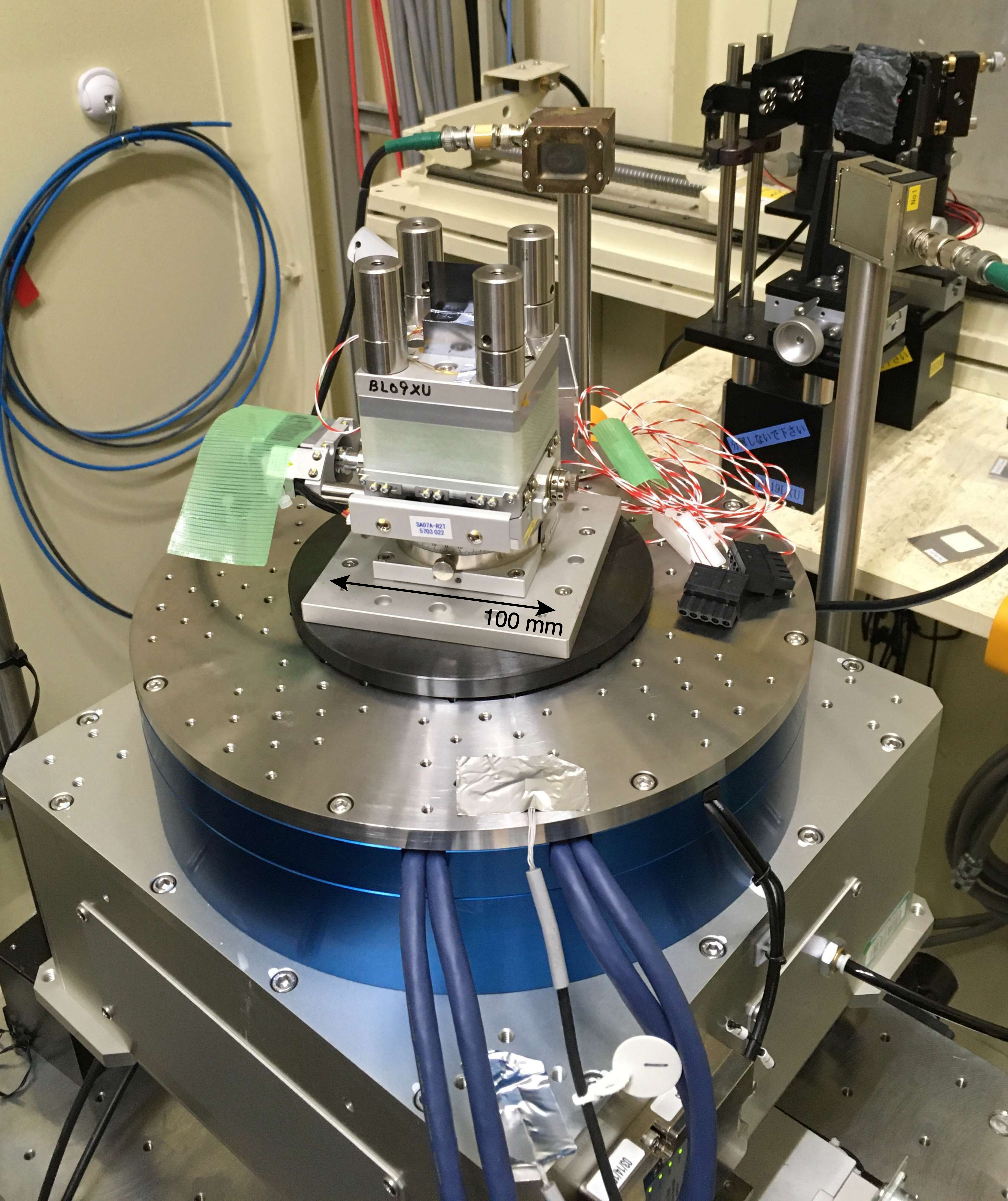}
\caption{Photograph of Bond diffractometer. The black circular plate is the rotation table based on SelfA.}
\label{fig:selfapic}
\end{figure}

The Bond diffractometer comprises a rotation table based on SelfA, a manual rotation table, a Si plate, two motorized swivel stages,  and two silicon PIN photodiodes. Its schematic is shown in figure~\ref{fig:system}b, and a photograph is shown in figure~\ref{fig:selfapic}.

The Si plate was placed at the top of the apparatus. The thickness of the plate was \SI{0.5}{mm}. The Si plate was cut from the same ingot of the standard reference crystal used in a previous study \cite{Cavagnero2004,Cavagnero2004e} with natural isotopic compositions; in that study, the lattice spacing \(d_{220}\) of the (220) lattice planes was measured carefully. The Si plate was glued on an aluminum plate and was covered by an aluminum cover (not shown in figure~\ref{fig:system}), which had polyimide beam windows at both ends. It was also covered by foamed polystyrene to stabilize the temperature around the Si plate.
A glass epoxy G10 block was inserted between the crystal and the motorized components at the bottom for thermal isolation.
Two temperature sensors, calibrated to an absolute accuracy of greater than \(\Delta T = \pm 5\,\mathrm{mK}\) were set inside the aluminum cover in advance. The sensors were read by a thermometer readout module (Fluke Black Stack thermometer 1560 and Platinum Resistance Thermometer scanner 2562) with an absolute accuracy of \(\Delta T = \pm 10\,\mathrm{mK}\).

The Si plate was mounted on a stack consisting of a top motorized swivel stage, manual rotation table, rotation table based on SelfA, and bottom motorized swivel stage, from top to bottom.
The two swivel stages were used to align the mutual angles along the three axes.
The manual rotation table was used to check the uniformity of SelfA; the details are described in section \ref{sec:uniformity}. 

The rotation table based on SelfA was manufactured by e-motion system, Inc. All the electronics were housed in the bottom case. The rotation table was supported by an air-bearing.

The two silicon PIN photodiodes were used for detecting the diffracted beam. These photodiodes were placed such that the angle difference between the primary beam and the line from the Si plate to a photodiode was twice the Bragg angle. The sensitive area of the photodiodes was \( 28 \times 28 \, \mathrm{mm}^2\), and the thickness was \SI{500}{\micro m}. These were disposed at an angle of \SI{\sim45}{\degree} to the incident beam. The output current from the photodiodes was read out by current amplifiers (KEITHLEY 428).

\subsection{Measurement procedure} \label{sec:procedure}

\begin{figure}
\includegraphics[width=8.5cm, bb=0 0 566 385, clip]{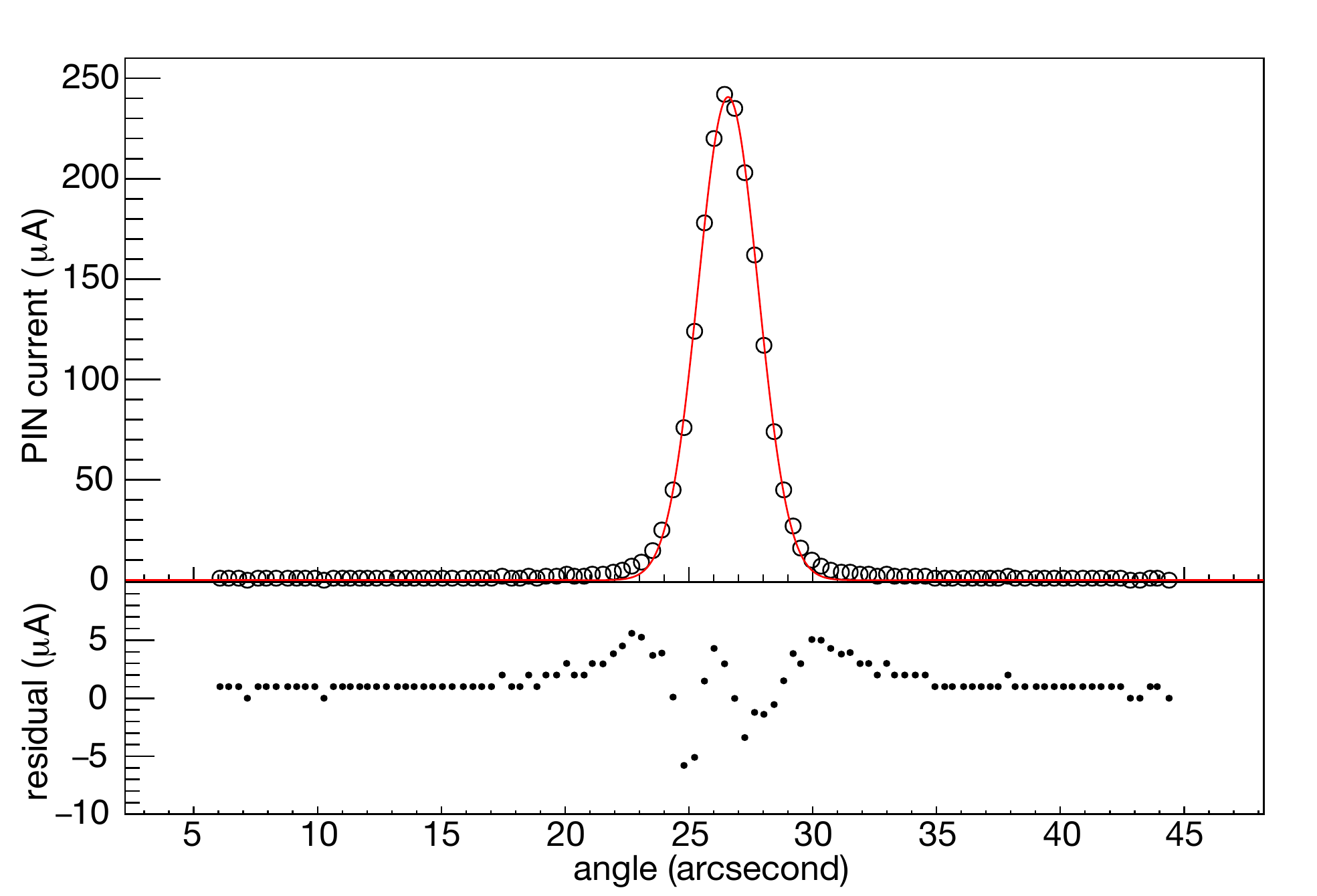}
\caption{(Top) X-ray diffraction peak measured on one side. The horizontal axis is the Si plate rotation angle controlled by SelfA, and the vertical axis is the output current from the PIN photodiode. (Bottom) The residual plot of the upper figure.}
\label{fig:diffractionpeaks}
\end{figure}

 For absolute energy measurement, the X-ray beam diffractions were monitored while rotating the Si plate. The X-ray diffraction occurred only when the angle between the primary beam and the reciprocal lattice vector of the crystal coincided with the Bragg angle. Figure~\ref{fig:diffractionpeaks} shows the diffraction peak monitored by one PIN photodiode. We rotated the rotary encoder by $\sim$0.4\,arcsecond and fixed it each time. We repeated this 0.4\,arcsecond rotation step 100 times as shown in the figure. It took $\sim$1 second for a step. The rotation angle at the diffraction center at each side of the primary beam was obtained by fitting with a Gaussian function. The angle difference between the two peaks \(\Delta \theta\) was \(2\theta_\mathrm{B}\). 
 Finally, the X-ray photon energy was determined by substituting the obtained \(\theta_\mathrm{B}\) into Eq.~\ref{eq:conversion}.
  
 Before the actual photon energy measurement, the two swivel stages should be adjusted. For this, we measured and minimized the angle difference \(\Delta \theta\) by rotating the swivel stages. We iterated the adjustment of the swivel stages individually. Next, we disconnected the power cable of the upper swivel stage, and we performed the self-calibration of SelfA. 
Since \SI{360}{\degree} rotation is required for SelfA calibration, the temperature sensor cables were also disconnected to prevent tangling.

\section{Measurement}

\subsection{NRS measurement}
\label{sec:nrs}

We performed the NRS measurement at the SPring-8 BL19LXU beamline \cite{Yabashi2001}.
We measured the resonance energy of the first excited state of the \(^{40}\)K nuclide, with a resonance energy and half-life of \(E_\mathrm{res} = 29829.9(6)\,\mathrm{eV}\) \cite{toi} and 4.24(9)\,ns \cite{Endt1990}, respectively. Its energy linewidth is sufficiently narrow compared with the energy bandwidth of the X-ray beam and can be ignored; thus, this excited state is a good target for the NRS measurement. We carried out the measurements in three beam times with two units of SelfA to verify reproducibility. The parameters are listed in Table~\ref{tab:runlist}.

In beam times 1 and 2, the bunch mode was A, in which 203 identical electron bunches were equally spaced with a time interval of 23.6\,ns in the storage ring; in beam time 3, the bunch mode was D, in which 15\% of the total current was shared in five bunches with a time interval of 684.3\,ns, and the remaining 85\% was equally shared in successive bunch trains in 1/7 of the circumference with a repetition rate of 1.966\,ns \cite{SPring8}.

For the SelfA, the first one, which was used for beam times 1 and 2, was manufactured in 2007, and the second one, which was used for beam time 3, was manufactured in 2019. Other parts such as the Si plate, temperature sensors, and swivel stages were identical for all beam times.

The experimental overview is shown in figure~\ref{fig:system}c. The X-ray beam was monochromatized by two silicon monochromators. The first Si(111) monochromator was a high heat-load monochromator (HHM). After passing through the HHM, the intensity was \SI{8E13}{photons/s}, and the energy bandwidth was \SI{3.4}{eV} full width at half maximum. The second monochromator was a high energy resolution monochromator (HRM); we used Si(440) or Si(660) as the HRM. The beam size was defined to be approximately \(0.4 \times 0.4 ~\mathrm{mm^2}\) by the slit located after the HRM. The beam intensity was \SI{4E12} (\SI{1E12}) photons/s, and the full width at half maximum of the energy bandwidth was 0.26 (0.10) eV for Si(440) (Si(660)). The beam intensity was monitored by using an ionization chamber positioned after the slit. The monochromatized beam passed through the \(^{40}\)K target and then through the Bond diffractometer at the downstream end.

The apparatus of the NRS measurement was almost identical to that used in the previous work \cite{Yoshimi2018,Masuda2019}, only the target was different. The \(^{40}\)K target was a KCl pellet (diameter \SI{3}{mm}, thickness \SI{0.5}{mm}) prepared by pressing 5\,mg KCl powder with \(\sim\)2\,MPa. Because the natural abundance of \(^{40}\)K is only 0.01\%, we used 4\% enriched potassium. The pellet was covered with two \(20\times20\)\,mm\(^2\) MgF\(_2\) substrates. 
The scattered X-ray photons were detected by a dedicated energy-sensitive X-ray detector \cite{Masuda2017,Masuda2019nim}.

\begin{table}
\caption{Parameters of NRS measurements.}
\begin{tabular}{lcccl}      % Alignment for each cell: l=left, c=center, r=right
\hline
Beam time & Bunch mode & HRM & SelfA  \\ \hline
1 & A & Si(440), Si(660) & 1 \\
2 & A & Si(440), Si(660) & 1 \\
3 & D & Si(660) & 2 \\ \hline
\end{tabular}
\label{tab:runlist}
\end{table}

\begin{figure}
\includegraphics[width=8cm, bb=0 0 540 380, clip]{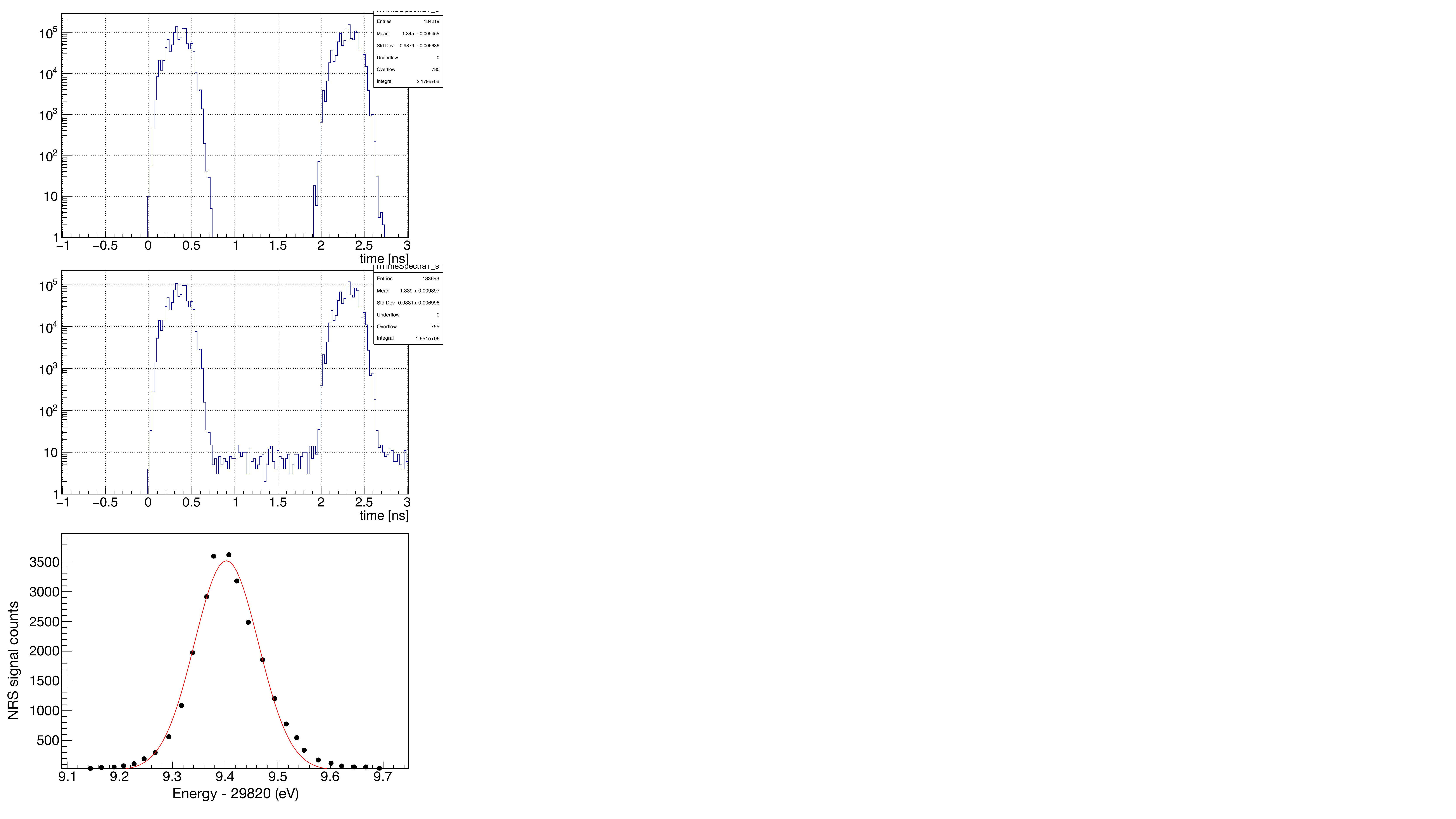}
\caption{Resonance curve of the first excited state of \(^{40}\)K. }
\label{fig:dmode}
\end{figure}

For the NRS measurements, the HRM was tuned to scan the photon energy. The typical resonance curve is shown in figure~\ref{fig:dmode}. 
We repeated the following procedure to obtain one resonance curve: rotate the HRM, measure the absolute energy of the photons, then accumulate the NRS data for \SI{100}{s}. One point of a resonance curve was obtained in 4--5 minutes. The resonance energy was obtained by fitting the resonance curve with a Gaussian function as shown in figure~\ref{fig:dmode}. The data points slightly deviated from the fitting function but no regular deviation pattern was found. The deviation could be caused by photon energy fluctuation or drifts over time because we took the data directly after rotating the HRM.
 
\subsection{Reproducibility and Uniformity}
\label{sec:uniformity}
\begin{figure}
\includegraphics[width=8.5cm, bb=20 0 950 700, clip]{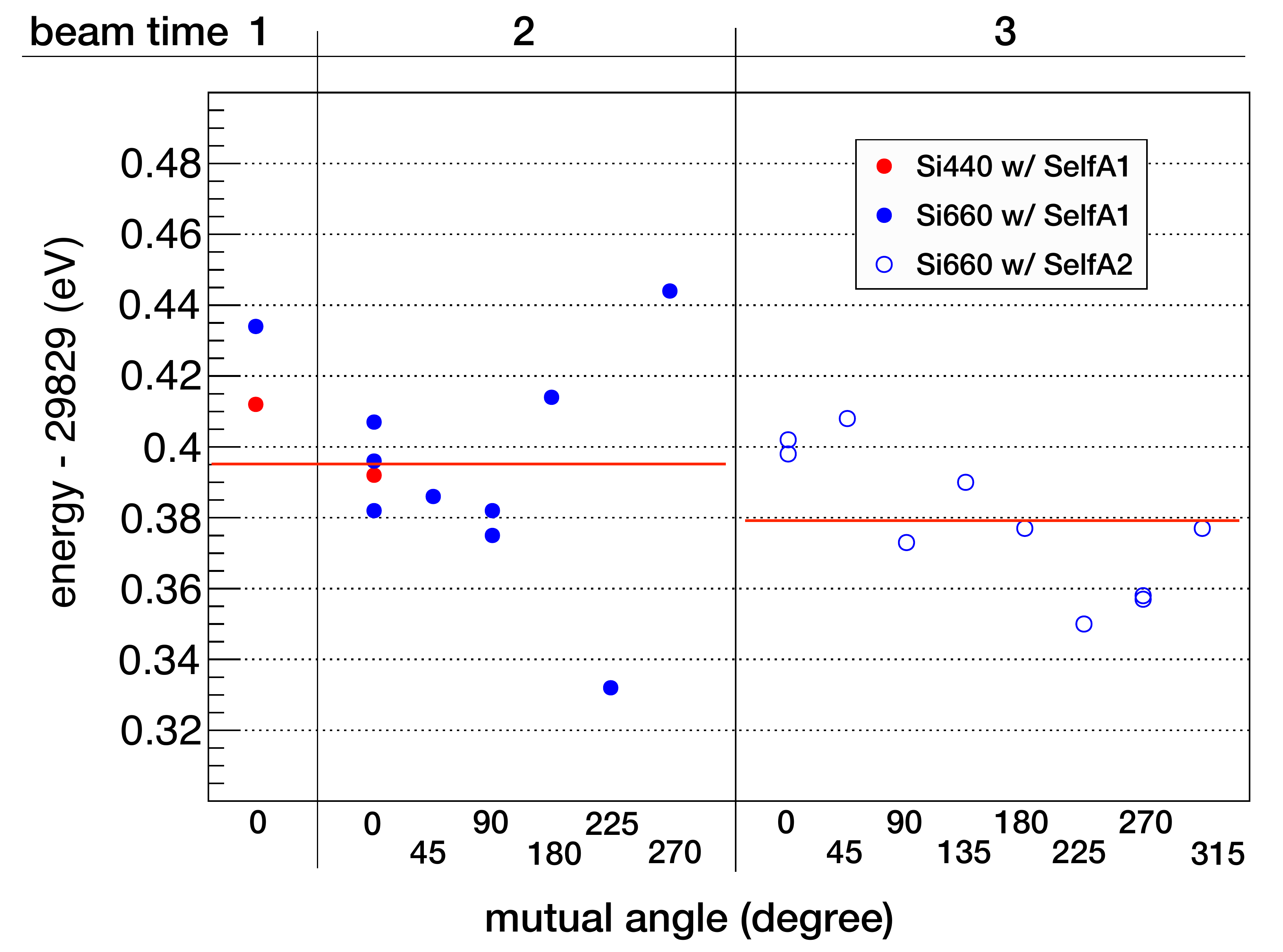}
\caption{Resonance peak value of each \(^{40}\)K NRS measurement. The numbers above indicate the beam time, and those at the bottom are the mutual angles between the Si plate and the rotation table. The horizontal red bars are the averaged values of the measurement with SelfA1 or SelfA2.}
\label{fig:40khistory}
\end{figure}

 To verify the reproducibility and uniformity of the system, we performed the NRS measurements for various mutual angles between the Si plate and the rotation table based on SelfA by rotating the manual rotation table between these. For beam times 2 and 3, the resonance peaks were measured with six and eight angle settings, respectively, as shown in figure~\ref{fig:40khistory}. We performed swivel adjustment and self-calibration function after every rotation of the manual rotation table.
The units of SelfA for beam times 1 and 2 were different from that for beam time 3; therefore, the mean value of the resonance energy was evaluated for each unit of SelfA. We found that the mean values of the resonance energy were consistent within \( \Delta E_\mathrm{res} = 0.02\,\mathrm{eV}\) (\( \Delta E_\mathrm{res}/E_\mathrm{res}=0.7\,\mathrm{ppm}\)) between the SelfA units, while the maximum deviation from the average was  \( \Delta E_\mathrm{res} = 0.06\,\mathrm{eV}\) (\( \Delta E_\mathrm{res}/E_\mathrm{res}=2.0\,\mathrm{ppm}\)). This demonstrates the reproducibility, including the individual difference in units of SelfA and beam time.

\subsection{Uncertainty analysis}
\subsubsection{Lattice spacing of the silicon crystal}
 We quoted the lattice spacing as \(d_{220} = 192.01559(2)\, \mathrm{pm}\), considering the inhomogeneity of the crystal \cite{Cavagnero2004,Cavagnero2004e}. The energy uncertainty due to this inhomogeneity was 0.10\,ppm.

For the thermal expansion effect, the uncertainty of the temperature and the thermal expansion coefficient were considered.
The difference between the two temperature sensors was \(\sim\)\SI{0.026}{K}. This difference can be considered as the temperature gradient inside the cover that arises because of the heat load from the lower motorized components. We considered half of the difference (\SI{0.013}{K}) as the uncertainty.
 A local temperature deviation inside the Si plate where it was actually irradiated by the X-ray beam will cause uncertainty.  The heat load of the Si plate due to the X-ray beam was estimated as \(\sim\)\SI{0.5}{mW}. We estimated that the local temperature increase at the irradiation spot was lower than \SI{0.01}{K} through a finite element analysis calculation.
The thermal expansion coefficient we used was \(C_\mathrm{temp}=\)~\SI{260.00E-8}{/K} based on a previous study \cite{Lyon1977}. 
Although the coefficient slightly depends on the temperature, it is sufficient to use the constant value because the average between the base temperature (\SI{22.5}{\celsius}) and the actual temperature (\SI{27.5}{\celsius}) is \(C_\mathrm{temp} = 260.11 \times 10^{-8}\,\mathrm{/K}\). Therefore, the temperature dependence of the coefficient was not taken into account. %The uncertainty of the energy due to the temperature and the thermal expansion coefficient are 0.04\,ppm and $<0.01$\,ppm, respectively.

For compression due to ambient pressure, we did not correct the pressure effect or use a pressure sensor in this study. Even if the uncertainty of the pressure was \( \Delta P/P = 10\%\), the uncertainty of the energy is less than 0.05 ppm because the compressibility is only \(C_\mathrm{comp} = 1.0221(3) \times 10^{-11}\,\mathrm{/Pa}\) \cite{Hall1967}; therefore, the compression effect was ignored in this study.

By summing the uncertainties mentioned above in quadrature, the uncertainty due to the lattice spacing was estimated to be \(\Delta d/d = 0.11\,\mathrm{ppm}\). These uncertainties and their contributions on the lattice spacing are summarized in table~\ref{tab:latticeerror}.

\begin{table*}
\caption{Uncertainty related to lattice spacing \(d\).\label{tab:latticeerror}}
\begin{tabular}{lccc}      % Alignment for each cell: l=left, c=center, r=right
\hline
Parameter & Value & Uncertainty & Contribution (ppm) \\ \hline
Lattice spacing $d_{220}$ & \SI{192.01559}{pm} & \SI{0.00002}{pm} & 0.10 \\
Temperature monitor $T$ & $\sim$\SI{27.5}{\celsius} & \SI{0.013}{K} & 0.03 \\
Local heating $T$ & -& \SI{0.009}{K} & 0.02 \\
Temperature expansion coefficient $C_\mathrm{temp}$ & \SI{260.00E-8}{/K} & \SI{0.11E-8}{/K} & $<$0.01 \\ \hline
Total & & & 0.11 \\
\hline
\end{tabular}
\end{table*}

\subsubsection{SelfA angle measurement}
 The uncertainty of the angle measurement by SelfA was classified into four parts.
 
The rotary encoder was servo controlled with a resolution of 0.035\,arcsecond per pulse. The quantization uncertainty of
\(0.035/2\sqrt{3}\)\,arcsecond \(\sim\) \,\SI{\pm0.010}{arcsecond} should be considered.

While the crystal angle was fixed, the angle deviated for one pulse. Therefore, the maximum angle deviation should be less than \(1\pm 1\) pulses, and the quantization uncertainty of a rectangular distribution with a width of 3 pulses (\(0.035\times 3 / 2\sqrt{3}\)\,arcsecond = \SI{\pm0.030}{arcsecond}) was added to the uncertainty.

\begin{figure}
\includegraphics[width=8cm, bb=0 0 535 350, clip]{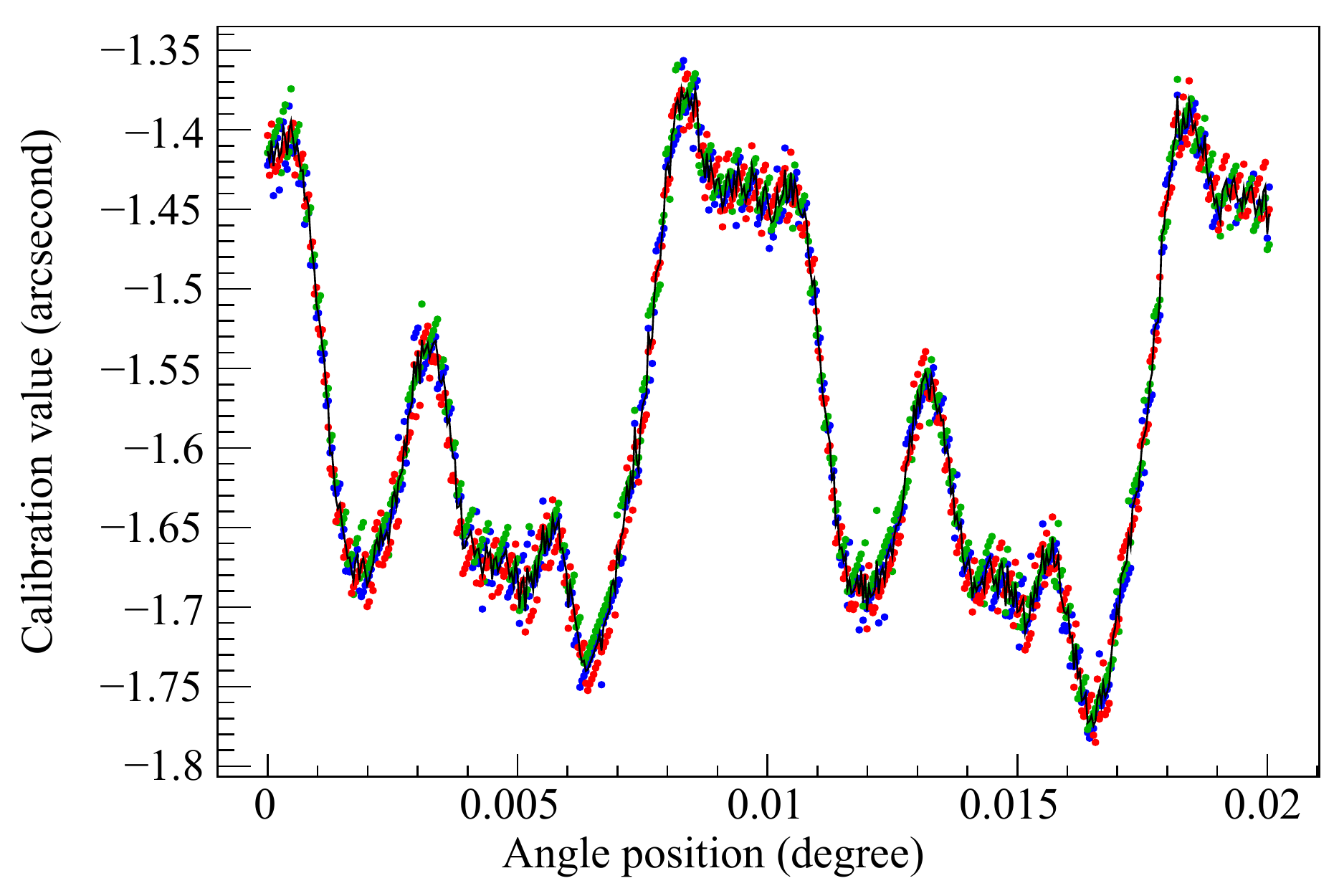}
\caption{Calibration values obtained by performing the self calibration function three times. The horizontal axis shows the rotation angle of the SelfA calculated based on the number of steps of the rotary encoder, and the vertical axis is the calibration value for each rotation angle. The colored points are individual calibration values; each color represents a different calibration sequence. The black line is the average of the three sequences.}
\label{fig:selfacalibsub}
\end{figure}

\begin{figure}
\includegraphics[width=8cm, bb=0 0 620 440, clip]{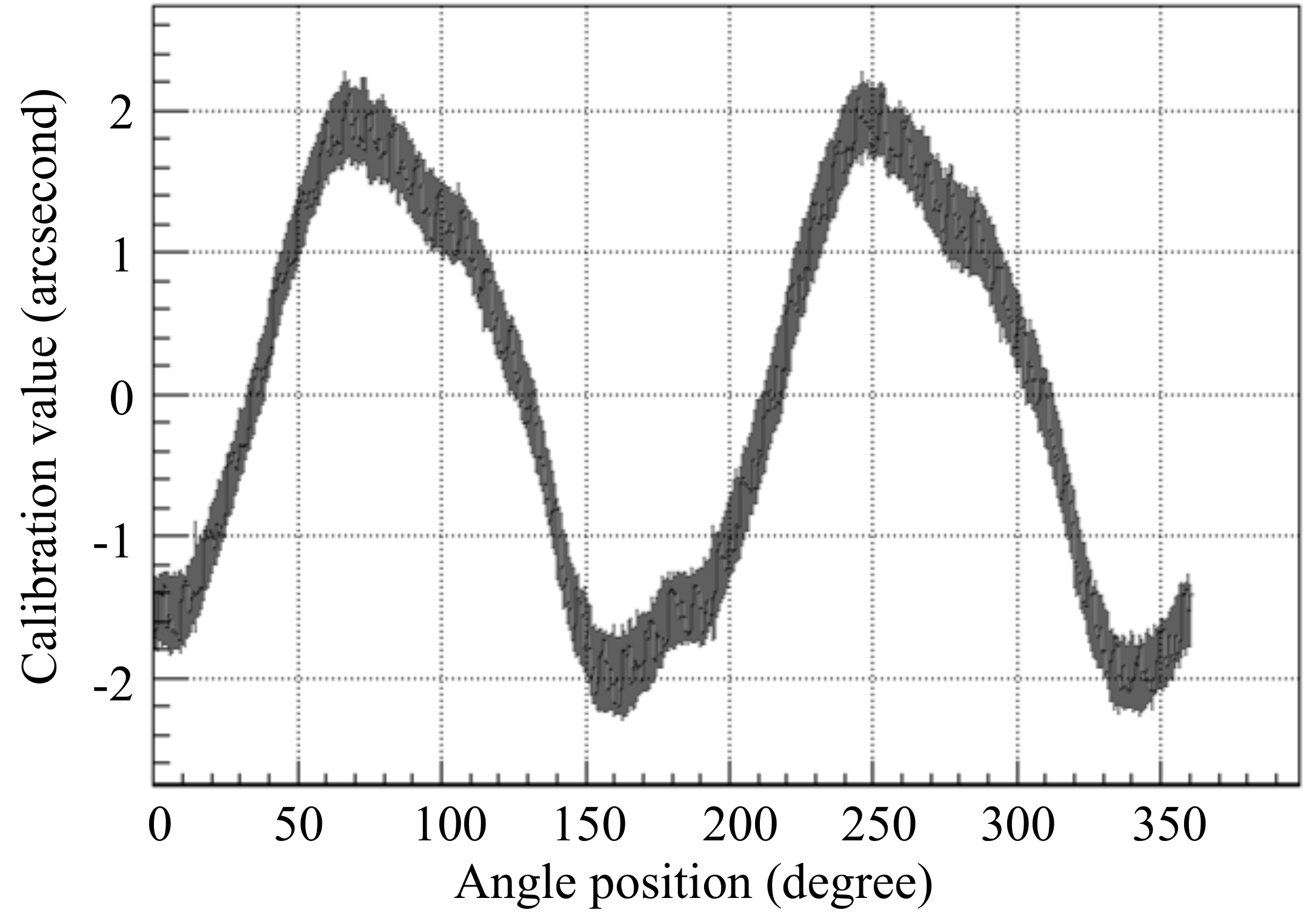}
\caption{Averaged calibration value for each angle position along the entire circumference of the rotary encoder.}
\label{fig:selfacaliball}
\end{figure}

The calibration values of SelfA could fluctuate at a certain level because of the variation in the measuring environment. To estimate the fluctuation, we performed three calibrations for each calibration sequence, and we used the averaged values of these three cycles. Figure~\ref{fig:selfacalibsub} shows the individual calibration values of the three cycles along with their average. Note that the figure is an enlarged view that shows only the \(0.02^\circ\) region from the entire \(360^\circ\) circumference.
Figure~\ref{fig:selfacaliball} shows the averaged calibration values for the entire circumference. We evaluated the standard deviation from the average to be less than \SI{\pm0.030}{arcsecond}.

\begin{figure}
\includegraphics[width=8cm, bb=0 0 380 250, clip]{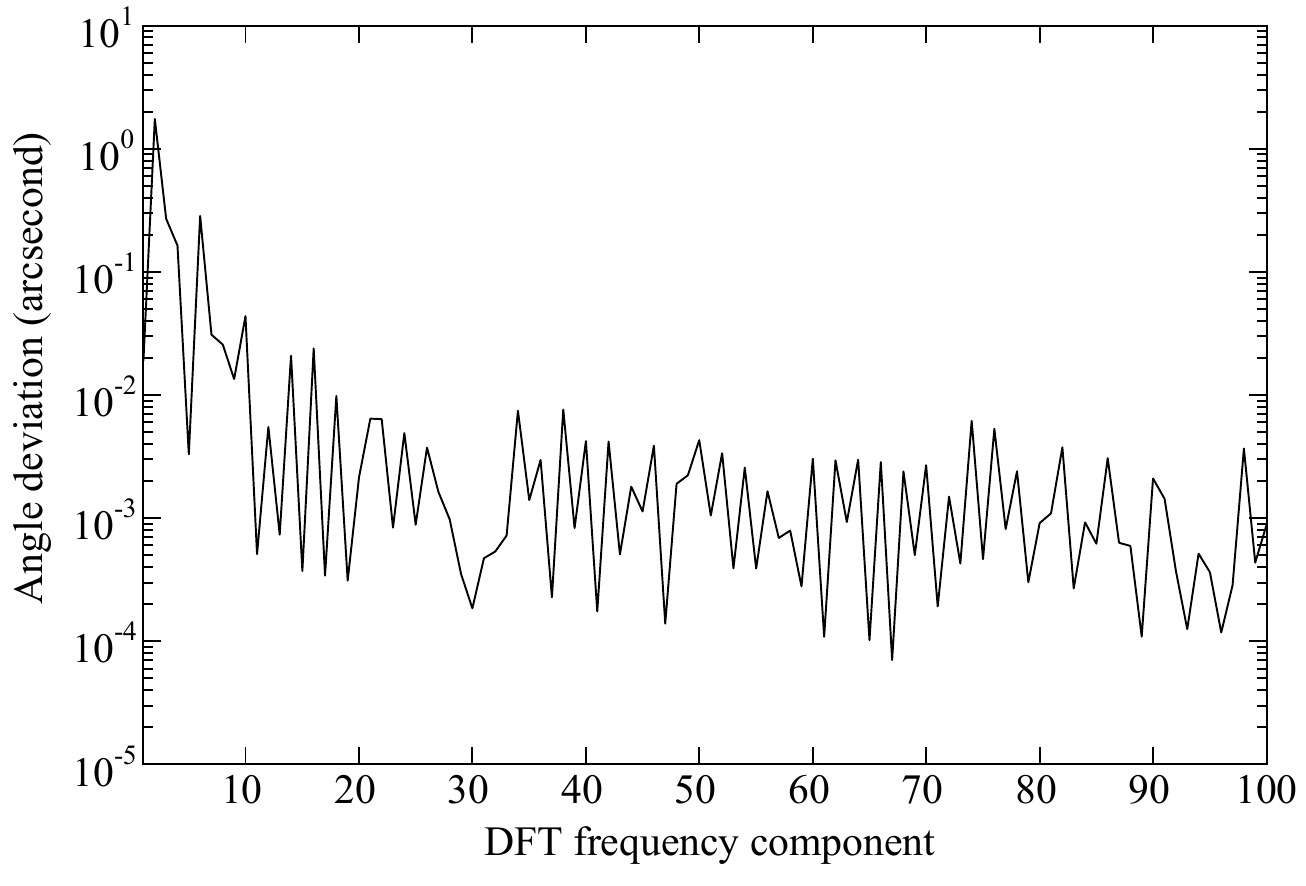}
\caption{DFT component below the 100th order.}
\label{fig:fftlower}
\end{figure}

\begin{figure}
\includegraphics[width=9cm, bb=0 0 400 250, clip]{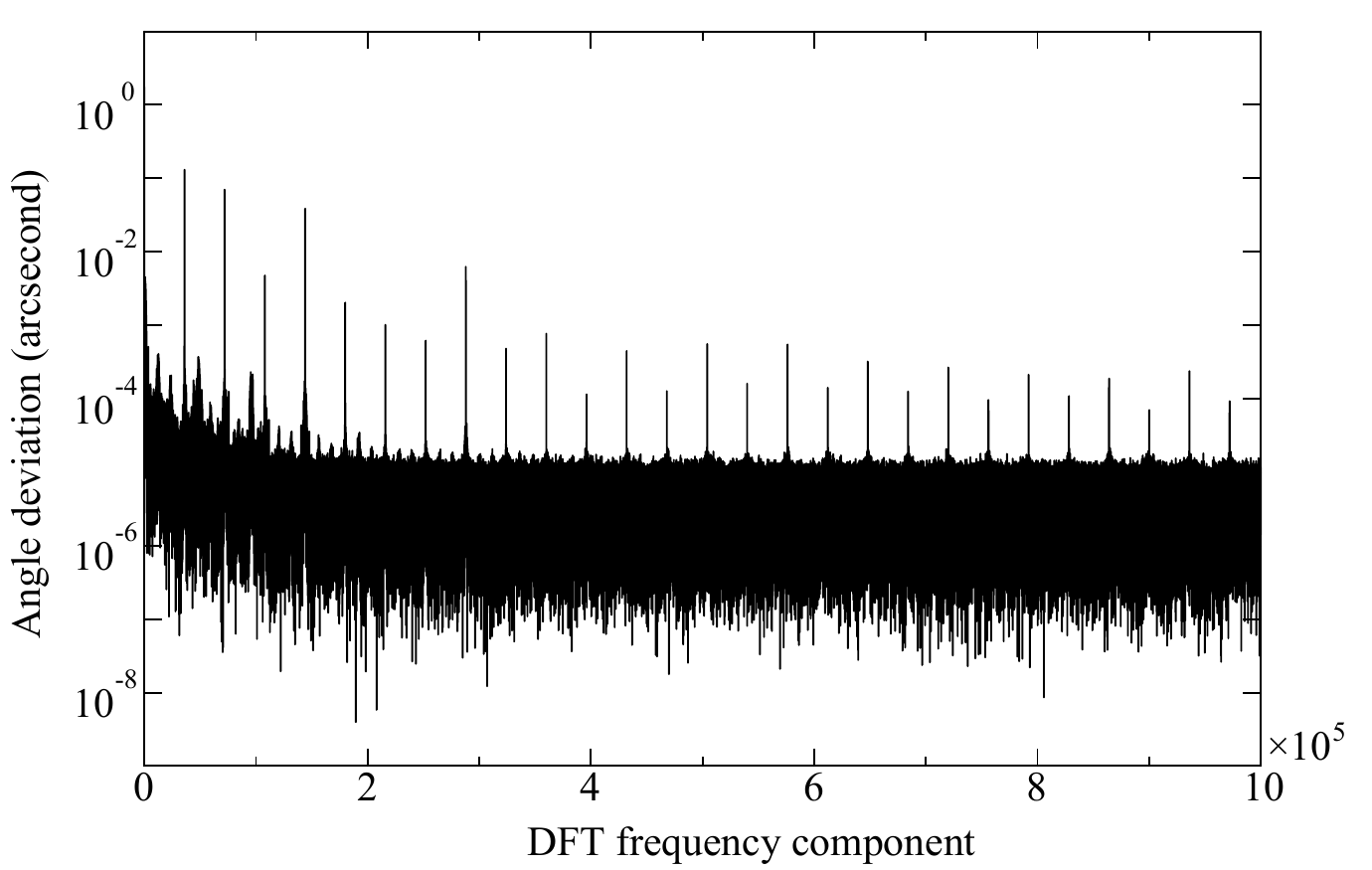}
\caption{DFT component up to the \(1\times10^6\)th order.}
\label{fig:ffthigher}
\end{figure}

The last factor was the higher order of the Fourier components that SelfA cannot detect. A detailed explanation on this analysis procedure is provided in a previous work \cite{Watanabe2014}. 
As mentioned in section \ref{sec:selfa}, the self-calibration function cannot detect the \(84n\)-th order Fourier components.
This magnitude can be estimated from the lower order components. Figures~\ref{fig:fftlower} and \ref{fig:ffthigher} show the typical discrete Fourier transformed (DFT) components of the calibration values. According to figure~\ref{fig:fftlower}, any components higher than the 18th order were less than \SI{0.01}{arcsecond}; thus, the \(84n\)-th order components could be assumed to be lower than \SI{0.01}{arcsecond} as well. 

The peak components in figure~\ref{fig:ffthigher} were due to the electrical interpolation, which appeared every 36000 components. The \(7n\)-th order components of the interpolation (252000th) cannot be detected, and these can be assumed to be less than \SI{0.01}{arcsecond}. 
 
By summing the uncertainties listed in this subsection in quadrature, the angle uncertainty due to the SelfA was estimated to be \SI{0.044}{arcsecond}, and the corresponding photon energy uncertainty was \( \Delta E_\mathrm{Xray} / E_\mathrm{Xray} = 0.67\,\mathrm{ppm}\). This was the dominant uncertainty in our estimations. These uncertainties in the angle measurement are listed in table~\ref{tab:selfaerror}.

\begin{table*}
\caption{Uncertainty related to angle measurement.\label{tab:selfaerror}}
\begin{tabular}{lcc}      % Alignment for each cell: l=left, c=center, r=right
\hline
Parameter & Value & Angle uncertainty  \\ \hline
Angle quantization & 0.035\,arcsecond & \SI{0.010}{arcsecond} \\
Servo control quantization & 0.105\,arcsecond & \SI{0.030}{arcsecond} \\
Calibration fluctuation & & \SI{0.030}{arcsecond} \\
Higher-order Fourier components & & $<$\SI{0.01}{arcsecond} \\ \hline
Total & & \SI{0.044}{arcsecond} \\
\hline
\end{tabular}
\end{table*}

\subsubsection{Other sources}
 The effects due to the uncertainty of the swivel setting were estimated from the repeatability of the angle measurement of the swivels. The repeatability was conservatively \(\Delta \theta_\mathrm{beam} = \Delta \theta_\mathrm{recip} = 0.01^\circ\), and the corresponding photon energy uncertainty was 0.03\,ppm.
Because the diffraction peak shapes (shown in figure \ref{fig:diffractionpeaks}) are dominated by the horizontal angular divergence of the incident beam, the shapes on both sides are supposed to be identical; therefore, the difference between these sides should be considered as the systematic uncertainty. We estimated this effect by fitting the shapes to various functions as well as to the simple Gaussian function, and we conservatively adopted the maximum deviated function. The resulting uncertainty was \SI{0.0083}{arcsecond} for $\theta_\mathrm{B}$; the corresponding photon energy uncertainty was 0.17\,ppm.
 The statistical precision of the diffraction peak center determination by the Gaussian fit was estimated as \SI{0.0095}{arcsecond} for a diffraction peak. The corresponding photon energy uncertainty was 0.14\,ppm.
The last uncertainty is categorized as a random error, while the other uncertainties are systematic for the X-ray photon energy measurement because it is reasonable to suppose that the swivels setting and the beam angle divergence are unchanged during measurements.
 
In addition, we checked the dependence of the rotation direction on the rotation table and beam intensity, which were not supposed to affect the measurements. The change in the measured resonance energy was 0.23\,ppm when the rotation direction was reversed.
 For the beam intensity effect, we reduced the beam intensity to half, and the measured resonance energy was stable within 0.04\,ppm from the original intensity. 
Since these two effects were smaller than the estimated uncertainty, they can be ignored.

\begin{table}
\caption{Energy uncertainty estimation list.\label{tab:error}}
\begin{tabular}{lc}      % Alignment for each cell: l=left, c=center, r=right
\hline
Parameter & Contribution (ppm) \\ \hline
Lattice spacing & 0.11 \\
Angle measurement by SelfA & 0.67 \\
Diffraction center determination$^*$ & 0.14 \\
Diffraction peak shape & 0.17 \\
Swivel setting & 0.03 \\
\hline
{\footnotesize $^*$random}
\end{tabular}
\end{table}

\subsection{Results}
 Among the NRS measurements for the three beam times, the maximum deviation from the average of the measured resonance energy was observed at a level of 2\,ppm, while the estimated uncertainty of the photon energy measurement, which we discussed in this section, was 0.7\,ppm as a standard deviation in total. Table~\ref{tab:error} summarizes the uncertainties discussed in this section. 
 
We determined the excitation energy by averaging all the values, and we conservatively quoted the maximum deviation as the uncertainty: \SI{29829.39\pm0.06}{eV}. This result is one order of magnitude better than the previously reported value of \SI{29829.9\pm0.6}{eV} \cite{toi}.

\section{Conclusion}
 We reported a new absolute X-ray energy measurement method. The method uses a Bond diffractometer with a silicon single crystal plate and a commercially available rotation table. We measured the resonance energy of the first excited state of the \(^{40}\)K nuclide via the NRS technique and demonstrated the performance of the proposed method. The results, which were obtained using two different units of SelfA, showed a good agreement, better than \(\Delta E/E = 0.7\,\mathrm{ppm}\). While the estimated uncertainty of the photon energy measurement is at a level of  0.7\,ppm, the observed reproducibility was found to be 2\,ppm at the maximum deviation from the average. 
We improved the energy measurement of the first excited nuclear state of the \(^{40}\)K nuclide by one order of magnitude. This improvement is achieved based primarily on the high-accuracy angle encoder, SelfA, which is able to determine the rotation angle with an accuracy on the order of 0.1\,arcsecond.
The system can be applied to a wide energy range of the X-ray beam and enables fast and easy \textit{in-situ} photon energy calibration.

    % Appendices appear after the main body of the text. They are prefixed by
     % a single \appendix declaration, and are then structured just like the
     % body text.

\section*{Acknowledgements}
The synchrotron radiation experiments were performed at the BL19LXU line of SPring-8 with the approval of the Japan Synchrotron Radiation Research Institute (JASRI) (proposals 2018A1326 and 2018B1436) and RIKEN (proposal numbers 20180045 and 20190051). We thank all the staff at SPring-8, especially T. Kobayashi and K. Ishino, for building the operating software of the Bond diffractometer and for technical support for the experiment. We would like to thank Editage (www.editage.com) for English language editing. This work was supported by JSPS KAKENHI grants JP18H01230, JP19H00685, JP19K14740, and JP19K21879. This project has received funding from the European Research Council (ERC) under the European Union's Horizon 2020 research and innovation programme (Grant agreement No. 856415 ``ThoriumNuclearClock''). T.M. acknowledges the INAMORI foundation.

\appendix
\setcounter{figure}{0}

\section{Reproducibility with fixed photon energy}
In this section, we describe the stability and uniformity measurements with fixed photon energy instead of the NRS measurement. These measurements were done during beam time 3.  The angle of both HHM and HRM were fixed, and the photon energy measurement was repeated for various mutual angles.
As we described in the introduction, the absolute photon energy can easily deviate because of changes in the thermal distribution of monochromators. This cannot be avoided in the NRS measurement because the photon energy has to be scanned. The advantage of this study is that the monochromators can be fixed in the beam line. 
However, this measurement cannot distinguish between the fluctuation of the photon energy itself and that of the measurement system; therefore, it only indicates an upper bound in a short time range without the energy scan.

\begin{figure}[h]
\includegraphics[width=8cm, bb=0 0 550 350, clip]{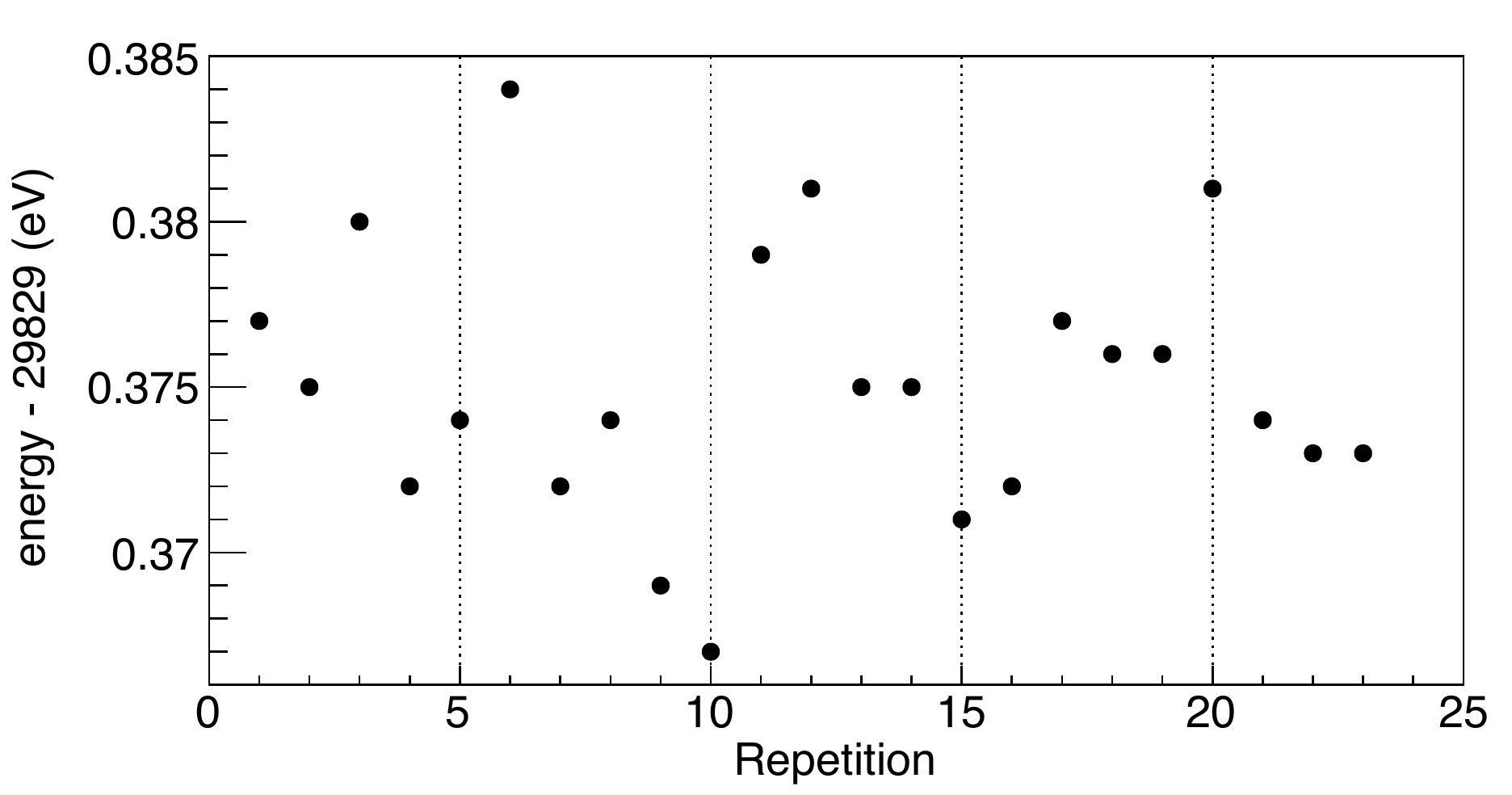}
\caption{Repeated photon energy measurements.}
\label{fig:stability_fixed}
\end{figure}

\begin{figure}[h]
\includegraphics[width=9cm, bb=0 0 540 250, clip]{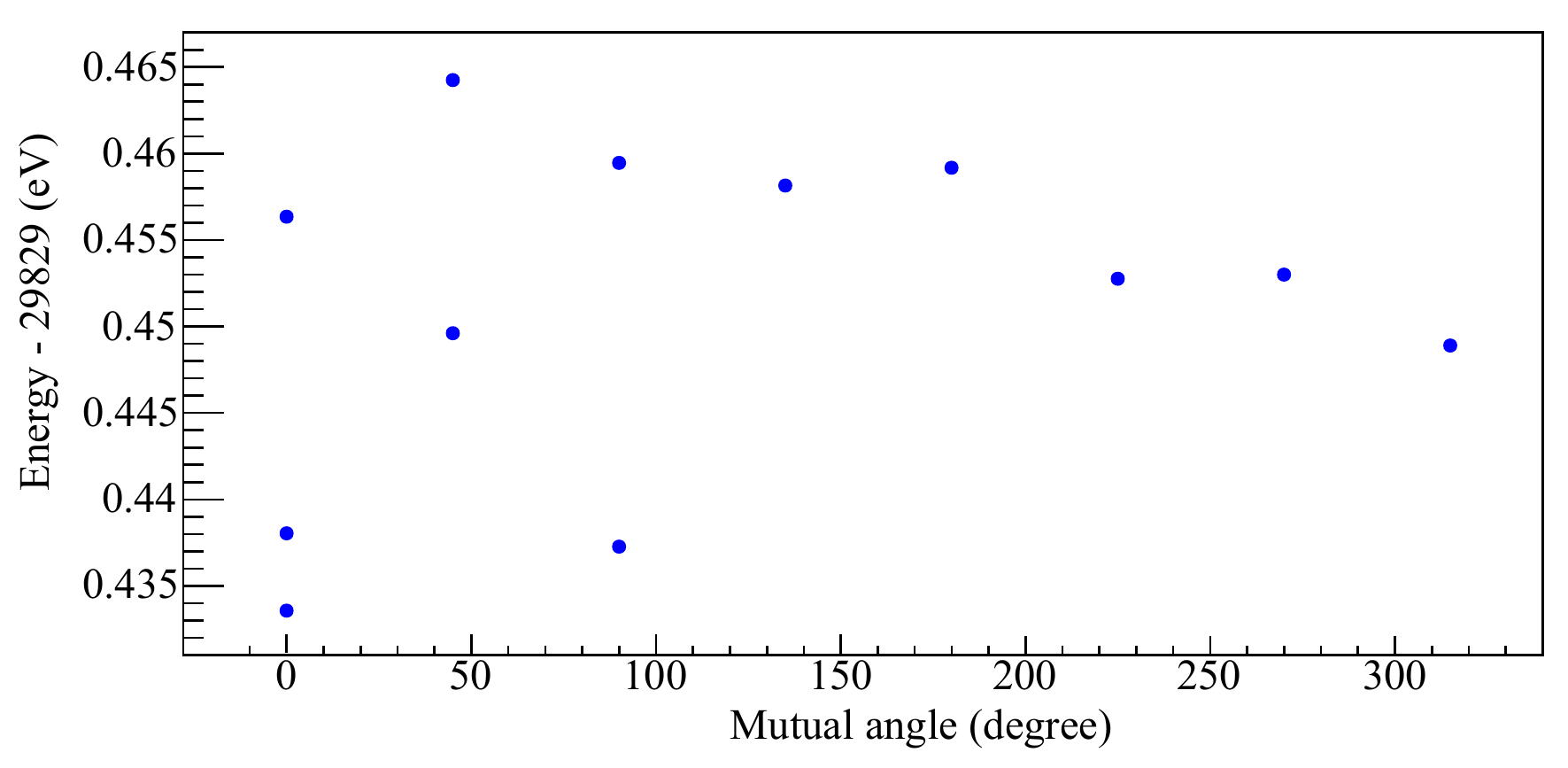}
\caption{Photon energy measurement at various angle positions of SelfA.}
\label{fig:uniformity_fixed}
\end{figure}

Figure~\ref{fig:stability_fixed} shows the results of the energy, which was measured 23 times in \(\sim\)85 minutes. The manual rotation table was fixed during the run. The peak-to-peak deviation was \SI{17}{meV}. 
Figure~\ref{fig:uniformity_fixed} shows the results of the measured energy for various mutual angles. The time required was 10--15 minutes for one point and \(\sim\)150 minutes for all points. The peak-to-peak deviation was \SI{30}{meV}; this deviation is smaller than the observed uncertainty described in the main text.

     %-------------------------------------------------------------------------
     % The back matter of the paper - acknowledgements and references
     %-------------------------------------------------------------------------

     % Acknowledgements come after the appendices

\bibliographystyle{unsrt}
\bibliography{Bond}

\end{document}